\documentclass[pra,aps,twocolumn,superscriptaddress,nofootinbib,showpacs]{revtex4}

\usepackage{latexsym}
\usepackage{hyperref}
\usepackage{epsfig}
\usepackage{graphics}
\usepackage{amsmath}
\usepackage{xspace}
\usepackage{amssymb}
\usepackage{amsthm}
\theoremstyle{plain}
\newtheorem{theorem}{Theorem}
\newtheorem{lemma}{Lemma}
\newtheorem{corollary}{Corollary}
\theoremstyle{definition}
\newtheorem{defn}{Definition}

\theoremstyle{remark}

\begin{document}

\title{Fungible dynamics: there are only two types of entangling
  multiple-qubit interactions}

\author{Michael J. Bremner}
\email[]{bremner@physics.uq.edu.au}
\author{Jennifer L. Dodd}
\email[]{jdodd@physics.uq.edu.au}
\affiliation{School of Physical Sciences, \\
The University of Queensland, QLD 4072, Australia}
\affiliation{Centre for Quantum Computer Technology, \\
The University of Queensland, QLD 4072, Australia}
\affiliation{Institute for Quantum Information, California Institute
of Technology, Pasadena CA 91125, USA}

\author{Michael A. Nielsen}
\email[]{nielsen@physics.uq.edu.au and www.qinfo.org/people/nielsen}
\affiliation{School of Physical Sciences, \\
The University of Queensland, QLD 4072, Australia}
\affiliation{School of Information Technology and Electrical Engineering, \\
The University of Queensland, QLD 4072, Australia}
\affiliation{Institute for Quantum Information, California Institute
of Technology, Pasadena CA 91125, USA}

\author{Dave Bacon}
\email[]{dabacon@cs.caltech.edu}
\affiliation{Institute for Quantum Information, California Institute
of Technology, Pasadena CA 91125, USA}
\affiliation{Department of Physics, California Institute of
Technology, Pasadena CA 91125, USA}

\date{\today}

\begin{abstract}
  What interactions are sufficient to simulate arbitrary quantum
  dynamics in a composite quantum system?  It has been shown that all
  two-body Hamiltonian evolutions can be simulated using \emph{any}
  fixed two-body entangling $n$-qubit Hamiltonian and fast local
  unitaries.  By \emph{entangling} we mean that every qubit is coupled
  to every other qubit, if not directly, then indirectly via
  intermediate qubits.  We extend this study to the case where
  interactions may involve more than two qubits at a time.  We find
  necessary and sufficient conditions for an arbitrary $n$-qubit
  Hamiltonian to be \emph{dynamically universal}, that is, able to
  simulate any other Hamiltonian acting on $n$ qubits, possibly in an
  inefficient manner.  We prove that an entangling Hamiltonian is
  dynamically universal if and only if it contains at least one
  coupling term involving an \emph{even} number of interacting qubits.
  For \emph{odd} entangling Hamiltonians, i.e., Hamiltonians with
  couplings that involve only an odd number of qubits, we prove that
  dynamic universality is possible on an encoded set of $n-1$ logical
  qubits.  We further prove that an odd entangling Hamiltonian can
  simulate any other odd Hamiltonian and classify the algebras that
  such Hamiltonians generate.  Thus, our results show that up to local
  unitary operations, there are only two fundamentally different types
  of entangling Hamiltonian on $n$ qubits.  We also demonstrate that,
  provided the number of qubits directly coupled by the Hamiltonian is
  bounded above by a constant, our techniques can be made efficient.
\end{abstract}

\maketitle

%
%

\section{Introduction}
\label{sec:introduction}

One of the main goals of quantum information science is to
characterize the physical resources that are \emph{universal} for
quantum computation and simulation. Recently, the role of
entangling quantum dynamics has been studied in depth, and it has
been shown that a fixed two-body entangling Hamiltonian evolution
acting on a system, plus the ability to intersperse local unitary
operations, can be used to efficiently simulate any other two-body
Hamiltonian and hence is universal for quantum computation (see,
for
example~\cite{Dodd02a,Wocjan02a,Bennett01a,Leung01a,Dur01a,Nielsen02d,Wocjan02b,Wocjan02c,Vidal01b,Vidal01c},
and references therein.)

We extend this study to Hamiltonians containing interaction terms
involving more than two qubits at a time.  Using results from quantum
control theory, we determine what dynamics can be simulated with an
\emph{arbitrary} fixed $n$-qubit Hamiltonian, and complete local
control in the form of one-qubit unitary operations.  In particular,
we derive a simple criterion determining which Hamiltonians are
universal given local unitary operations.

In contrast to two-body entangling Hamiltonians, which are always
universal given local
operations~\cite{Dodd02a,Wocjan02a,Bennett01a,Leung01a,Dur01a,Nielsen02d,Wocjan02b,Wocjan02c,Vidal01b,Vidal01c},
we will see that not all many-body entangling Hamiltonians are
universal, given arbitrary one-qubit operations as the only
additional resource. This was previously noted for a specific
example by Vidal and Cirac~\cite{Vidal01b}.  Thus, unlike two-body
Hamiltonians, many-qubit entangling Hamiltonians are not all
equivalent up to local operations. The situation is somewhat
analogous to the study of multi-party entangled states, where it
has been shown that there exist different types of entanglement,
inequivalent up to local operations and classical
communication~\cite{Dur00b,Linden99a,Bennett01b} (LOCC). For
example, it is now well known that the ``GHZ'' and ``W'' states
are not equivalent up to LOCC.



Consider the following illustrative example. Suppose we are given
a Hamiltonian acting on three qubits, $H=X \otimes X \otimes I +
I\otimes X \otimes X$, where $X$ is the usual Pauli $\sigma_{x}$
operator. Then, if we can perform arbitrary local unitary
operations on each of the qubits, it has been shown in
\cite{Dodd02a,Wocjan02a,Bennett01a,Leung01a,Dur01a} that it is
possible to simulate any other Hamiltonian interaction on three
qubits, such as $H'= X\otimes X \otimes X$.  We call such a
Hamiltonian \emph{universal}. On the other hand, given $H'$ and
arbitrary local unitaries, it turns out that it is not possible to
simulate $H$, and thus $H'$ is not universal. A proof of this is
given in~\cite{Vidal01b}; both these results will also follow from
the general results given in this paper.

We address the problem of universality in full generality by giving a
necessary and sufficient condition for a Hamiltonian to be universal.
Our condition is dependent only on simple properties of the
Hamiltonian's decomposition into tensor products of Pauli operators.
The proof of this condition is constructive, in the sense that it
provides, in principle, an algorithm for using a universal Hamiltonian
to simulate any other interaction.  However, the techniques used in
our construction are not especially practical, especially in the
presence of noise, and it remains to be seen if more practical
constructions are possible.

%
%
In addition to our criterion for universality, we also examine what
can be done when a given Hamiltonian is not universal.  In particular,
we show that there always exists a simple encoding scheme to make
these Hamiltonians universal.

%
%
Let us make a more precise statement of our results.  Suppose we are
given a Hamiltonian $H$ acting on $n$ qubits, which can be written
uniquely in terms of its Pauli operator expansion
\begin{equation}
H =\sum_{j_1,\ldots,j_n=0}^3 h_{j_1 \ldots j_n}
\sigma_{j_1} \otimes \ldots \otimes \sigma_{j_n},
\end{equation}
where the $h_{j_1 \ldots j_n}$ are real numbers and
$\sigma_1,\sigma_2,\sigma_3$ are the Pauli sigma matrices $X, Y, Z$,
respectively, with $\sigma_0 \equiv I$ the identity.  We say that a
subset $S$ of the qubits is \emph{coupled} by this Hamiltonian if
there is a non-zero term in $H$ coupling those specific qubits.

%
%
When is $H$ universal?  An obvious condition is that the set of qubits
coupled by $H$ must be \emph{connected}.  That is, it should not be
possible to partition the qubits into non-trivial sets, $S$ and
$\overline S$, such that every term in the Pauli operator expansion
couples either a subset of $S$ or a subset of $\overline S$.  If this
were the case then $H$ could not be used to generate entanglement
between the qubits in $S$ and the qubits $\overline S$, and thus would
not be universal.  We say that a Hamiltonian connecting all the qubits
in this way is an \emph{entangling} Hamiltonian.  Note that this
definition may be restated in the language of graph theory: if qubits
correspond to vertices in a hypergraph, and couplings between qubits
correspond to hyper-edges, then the condition that the Hamiltonian is
entangling corresponds to the condition that the hypergraph is
connected.

%
%
With this background, our main results are easily stated.  Our first
result is that an entangling Hamiltonian and local unitaries is
universal if and only if the Pauli operator expansion for the
Hamiltonian contains a term coupling an even number of qubits.  This
result provides a simple, easily checkable criterion to determine
whether or not a Hamiltonian is universal. Returning to our previous
examples $H = X\otimes X \otimes I + I \otimes X \otimes X$ and $H' =
X \otimes X \otimes X$, this criterion tells us that $H$ is universal
when assisted by local unitaries, while $H'$ is not, in agreement with
the earlier claims.

%
%
Our second result concerns what happens when the Pauli operator
expansion contains only odd terms, and thus is not universal.  We say
that such a Hamiltonian is an \emph{odd} entangling Hamiltonian.  We
will prove that an odd entangling Hamiltonian acting on $n$ qubits is
capable of simulating any other odd Hamiltonian acting on those
qubits.  Thus, the odd entangling Hamiltonians are a fungible physical
resource, since having any one is equivalent to having any other, up
to local unitary operations. Furthermore, we show that an odd
Hamiltonian, together with local unitaries, generates either the
simple Lie algebra $\mathfrak{so}(2^n)$ or $\mathfrak{sp}(2^n)$,
depending on the number of qubits $n$ that are connected by the
Hamiltonians.

%
%
Our third result also concerns odd Hamiltonians.  We prove that, with
appropriate encoding, an odd entangling Hamiltonian and local
unitaries is universal on a set of $n-1$ logical qubits.  Thus, there
is not too great a loss in space efficiency when one attempts to use
such a Hamiltonian to simulate an arbitrary interaction.

%
%
Our results thus completely classify what can be achieved with an
$n$-qubit Hamiltonian and local unitary operations.  They demonstrate
that there are essentially only two types of Hamiltonians up to local
unitary operations: those whose Pauli operator expansion contains only
odd parity terms, and those with at least one even term.

%
%
An important caveat to our results concerns \emph{efficiency}.  When
we state that a set of interactions is \emph{universal} on a set of
qubits, we mean simply that those interactions can be used to simulate
any other interaction, without any claim as to whether the simulation
procedure is efficient, or otherwise.  We will say such a set of
interactions is \emph{dynamically universal}.  By contrast, in the
context of quantum computing, a set of resources is said to be
\emph{universal for quantum computation} if it can be used to simulate
a standard set of universal gates, such as the controlled-{\sc not}
and one-qubit unitary gates, \emph{with an overhead that is at most
polynomial in the number of qubits}.

%
%
Our results thus concern dynamic universality, and do not directly
address the question of universal quantum computation.  However, some
general observations may be made about the efficiency of our
constructions.  When the number of qubits that are directly coupled by
the Hamiltonian is bounded above by some constant $k$, then our
simulation techniques for gates such as the controlled-{\sc not} only
incur an overhead polynomial in the total number of qubits.  Thus, when
the coupling size is bounded, our results give criteria not only for
dynamic universality, but also for universal quantum computation.  By
contrast, when the number of qubits involved in couplings is
unbounded, our simulation technique is not polynomial, and thus our
results cannot be applied to deduce anything about universal quantum
computation.  Indeed, we conjecture that the two concepts of
universality do not, in general, coincide.

%
%
For the remainder of this paper, when we speak of a set of couplings
being universal, we mean dynamically universal, unless otherwise
stated.  The only exception is in Section~\ref{sec:efficiency}, which
contains the proof that when the terms in the Hamiltonian couple only
a bounded number of qubits, our techniques can be made efficient.

%
%
What is the significance of our findings?  It is tempting to conclude
that the main significance is for the design of quantum computers.
However, we do not believe our results are especially significant for
such questions.  Not only are our constructions impractical, but it is
an empirical fact that most interactions occurring in nature are
two-body interactions, and thus are adequately dealt with by earlier
work.

%
%
We believe our results are interesting for two other, less obvious,
reasons.  The first is the intrinsic interest in obtaining general
insights into quantum dynamics.  The fact that, given local unitary
control, there are only two different classes of Hamiltonian
evolution, seems to us a significant insight into the complicated
space of possible dynamical evolutions.  It tells us that dynamics
within each of these classes are \emph{fungible} physical resources.
In the case of two-party interactions this insight has led to the
beginnings of a quantitative theory of the
\emph{strength}~\cite{Nielsen02a,Childs03b} of dynamical operations,
much as the theory of entanglement dilution and
concentration~\cite{bennett96c} led to the quantitative theory of
entanglement~\cite{Bennett96a,Bennett96b}.

%
%
An interesting contrast is to the situation with state entanglement,
where the multipartite structure is complex and only partially
understood.  The number of classes of states which are inequivalent
under local operations and classical communication (LOCC) is enormous
--- for four-qubit states, there are already at least 9 inequivalent
classes~\cite{Verstraete02a}.  Our results thus demonstrate that there
are no direct analogies between multiparty state entanglement and
multiparty entangling dynamics --- in fact, the situation is
substantially simpler for dynamics than it is for states.

%
%
A second reason for interest is possible indirect applications.  For
example, although four-qubit interactions may not occur in nature, it
is certainly the case that interactions involving two objects with
four-dimensional state spaces may occur in nature.  Such systems can
naturally be mapped onto our problem by considering a single
four-dimensional system as being, effectively, a system of two qubits.
Constructions like this may make our results of interest, at least in
principle, for realistic physical systems.

The paper is structured as follows. Section~\ref{sec:background}
provides background and definitions.  Section~\ref{sec:proof}
establishes a body of general techniques for simulating one
Hamiltonian with another.  These techniques are applied in
Section~\ref{sec:universality} to prove our first main result, the
characterization of when a Hamiltonian is universal, assisted by
one-qubit unitaries.  Section~\ref{sec:odd} studies the non-universal
case, proving that any odd entangling Hamiltonian may be used to
simulate any other odd Hamiltonian.  We then describe an encoded
universality scheme that allows an odd entangling $n$-qubit
Hamiltonian to act universally on $n-1$ qubits, and provide a
Lie-algebraic classification of this case. In
Section~\ref{sec:efficiency} we show that our techniques can be made
efficient under certain conditions of bounded coupling size, and
finally in Section~\ref{sec:discussion} we summarize our results.

%
%

\section{Background and definitions}
\label{sec:background}

This section contains definitions and background material for the
remainder of the paper. We begin with some notation and definitions,
followed by a discussion of what it means to simulate one Hamiltonian
with another, and finally we review some previous work on the
universality of two-body Hamiltonians.

%
%

We now introduce some notational conventions. As stated in the
introduction, an arbitrary Hamiltonian $H$ on $n$ qubits can be
uniquely written in terms of Pauli operators via the Pauli operator
expansion
\begin{equation}
\label{eq:op_exp} H =\sum_{j_1,\ldots,j_n=0}^3 h_{j_1 \ldots j_n}
\sigma_{j_1} \otimes \ldots \otimes \sigma_{j_n},
\end{equation}
where the $h_{j_1 \ldots j_n}$ are real numbers and
$\sigma_1,\sigma_2,\sigma_3$ are the Pauli sigma matrices $X, Y, Z$,
respectively, with $\sigma_0 \equiv I$ the identity. Let the index
$\alpha$ denote each different combination $j_1 \dots j_n$
corresponding to non-zero terms in equation (\ref{eq:op_exp}). So for
each non-zero term in the Pauli expansion of $H$ we write $H_\alpha =
h_{j_1, ... ,j_n}\sigma_{j_1}\otimes ... \otimes \sigma_{j_n}$ and
thus $H=\sum_{\alpha} H_\alpha$.
\begin{defn}
  Let $C_\alpha$ denote the set of all Pauli product Hamiltonians that
  couple the same set of qubits as $H_\alpha = h_{j_1, ...
  ,j_n}\sigma_{j_1}\otimes ... \otimes \sigma_{j_n}$, that is
  $C_\alpha = \{\sigma_{k_1}\otimes ... \otimes\sigma_{k_n}| k_i = 0 \
  \text{iff\ } j_i=0 \}$. We call each set $C_\alpha$ a
  \emph{coupling} set.
\end{defn}
For example if $H_\alpha = X\otimes I \otimes Y\otimes Y$, then
$C_\alpha$ is the set of all products of Paulis acting non-trivially
on the same qubits, or $\{X \otimes I\otimes X\otimes X, X\otimes I
\otimes X\otimes Y,..., Z\otimes I \otimes Z\otimes Z\}$.
\begin{defn}
The set $S_\alpha$ is the set of qubits coupled by $H_\alpha$, or,
equivalently, all elements of $C_\alpha$.
\end{defn}
In the example above $S_\alpha= \{1,3,4\}$. Note that different
$H_\alpha$'s can give rise to the same $C_\alpha$ and $S_\alpha$.  We
will use the notation $|S_\alpha|$ to denote the number of qubits in
the set $S_\alpha$.

Now, what does it mean to simulate one Hamiltonian with another?  If
we can approximately induce evolution according to a Hamiltonian $H$
on an arbitrary state $|\psi\rangle$ for an arbitrary time $t$ without
actually using $H$, then we can simulate $H$, provided the
approximation can be made arbitrarily good.  This concept of
simulation is motivated by quantum computation which uses a universal
set of gates to simulate arbitrary unitary evolutions on a set of
qubits.

Our approach to the question of whether a Hamiltonian is universal is
to exhaustively build up the repertoire of different evolutions
simulatable with the Hamiltonian, in such a way that it becomes clear
whether or not the repertoire is a universal set of gates.

A first observation is that, given a term $H_\alpha=h_{j_1\cdots
j_n}\sigma_{j_1}\otimes\cdots\otimes\sigma_{j_n}$ we can simulate
$xH_\alpha$ for any real non-zero $x$ by adjusting the amount of time
we evolve according to $H_\alpha$.  Therefore, we can always simulate
$\tilde H_\alpha\equiv\pm\sigma_{j_1}\otimes\cdots\otimes\sigma_{j_n}$
with the sign given by the sign of $h_{j_1\cdots j_n}$.  If it is
negative, then we can make it positive by conjugating $\tilde
H_\alpha$ by a one-qubit unitary $U$ which anticommutes with it.
Thus, we can always obtain
$\sigma_{j_1}\otimes\cdots\otimes\sigma_{j_n}$.  For the remainder of
the paper, we will always assume that terms like $H_\alpha$ have this
form.

A simple, but important, observation is that, given the ability to
evolve according to some Hamiltonian $J$ and to perform a unitary
operation $U$ and its inverse $U^\dagger$, we can evolve according to
\begin{equation}
\label{eq:un_exp} Ue^{-iJt}U^{\dagger}=e^{-iUJU^\dagger t}.
\end{equation}
That is, we can simulate evolution according to the Hamiltonian
$J'=UJU^\dagger$.

%
%
This result can be used to show, for example, that given a Pauli
product Hamiltonian $H_{\alpha}$, we can simulate any other coupling
in $C_{\alpha}$, simply by performing local changes of basis on each
of the qubits to interchange the role of the $x, y$ and $z$ axes.
This is done by conjugating by one of the following three rotations:
\begin{equation}
e^{i\frac{\pi}{4}X},\ e^{i\frac{\pi}{4}Y},\ e^{i\frac{\pi}{4}Z}.
\end{equation}

Now, suppose we can evolve according to two Hamiltonians $J_1$ and
$J_2$.  Then for small times $\Delta$ the following identity holds
approximately:
\begin{equation}
\label{eq:ham_sum} e^{-iJ_1\Delta}e^{-iJ_2\Delta}\approx
e^{-i(J_1+J_2)\Delta}.
\end{equation}
That is, we can simulate evolution according to the Hamiltonian
$J_1+J_2$. Equation~(\ref{eq:ham_sum}) is important because it tells
us that if we are able to evolve a system according to a set of
different Hamiltonians, then it is possible to simulate arbitrary
linear combinations of elements of the set. We will treat this
identity as though it is exact for the remainder of the paper.  This
is justified for small $\Delta$. (See~\cite{Dodd02a} for an analysis
of the errors induced by this approximation, and the overhead required
to reduce them). The above identities are used extensively in this
paper, as they were in
\cite{Dodd02a,Nielsen02d,Wocjan02a,Wocjan02b,Wocjan02c,Bennett01a,Leung01a,Vidal01b,Vidal01c,Dur01a}.


It is also possible to simulate the commutator of two
Hamiltonians~\cite{Deutsch95a,Lloyd95a}, since
\begin{equation}
\label{eq:comm_sim}
e^{-iJ_1\Delta}e^{iJ_2\Delta}e^{iJ_1\Delta}e^{-iJ_2\Delta}\approx
e^{-i(i[J_1,J_2])\Delta^2}
\end{equation}
for small $\Delta$.  The error in this approximation is of order
$\Delta^3$, and can be made insignificant by choosing $\Delta$
sufficiently small.  This completes the basic set of tools that we use
to build up our repertoire of simulatable Hamiltonians. The reason for
this is that given a set of Hamiltonians $L= \{J_1,...,J_\zeta\}$ the
set of all simulatable Hamiltonians is given by the Lie algebra
generated by the set $L$~\cite{Wilcox67a} which can in turn be
generated with linear combinations and $i$ times commutators of
elements from $L$.

We can thus refine the central question of this paper to be: how does
the structure of the given Hamiltonian $H$ determine the Lie algebra
that can be generated by $H$ and arbitrary one-qubit evolutions?  This
is the question that we address in the remainder of this paper.

\section{Methods}
\label{sec:proof}
%
%
%

In the previous section we introduced some notation and basic tools.
This section is concerned with building up more sophisticated
simulation methods for the proofs of our main results, in later
sections.  In particular, there are two interesting simulation ideas
--- term isolation and commutator restriction --- that we will examine
in separate subsections.  These ideas may be more fully described as
follows:

\begin{itemize}

\item[\textbf{A.}] \textbf{Term isolation:}
  Given that we can simulate $H$ and perform
  arbitrary local unitaries, we can simulate an arbitrary term
  $H_\alpha$ in the expansion of $H$.  Recall that we write
  $H=\sum_\alpha H_\alpha$ where each $H_\alpha$ is a Pauli product
  Hamiltonian.

\item[\textbf{B.}] \textbf{Commutator restrictions:} We examine the
  restrictions placed on the simulation of commutators of coupling
  terms in $H$.
\end{itemize}

\subsection{Term isolation}
\label{subsec:termisolation}

In this section, we show that given $H$ we can simulate each term
$H_\alpha$ in $H$ using one-qubit unitaries and the composition
identities given in equations (\ref{eq:un_exp}) and
(\ref{eq:ham_sum}). Thus the capacity to simulate $H$ is equivalent up
to one-qubit unitaries to being able to simulate each $H_\alpha$.

For simplicity in the proof we now note that we can use one-qubit
unitaries to simulate the Hamiltonian $H^{(1)}= V_1\otimes \cdots
\otimes V_n H V_{1}^{\dagger}\otimes \cdots \otimes V_{n}^{\dagger}$
where the one-qubit unitaries $V_1,...,V_n$ are chosen so that the
term $H_\alpha$ has all $X$ and $Y$ operators in its expansion taken
to $Z$ and all $Z$ and $I$ operators left alone.  Thus in the
Hamiltonian $H^{(1)}$ every $\sigma_{j}\neq I$ in $H_\alpha$ is now
$Z$. Let us denote this term in $H^{(1)}$ by $H_{\alpha_Z}$. This
term, equivalent to $H_\alpha$, is the term that we wish to isolate.
>From now on we use the convention that the superscript on a simulated
Hamiltonian indicates a step in the algorithm, so $H^{(j)}$ would be
the Hamiltonian simulated after the $j^\text{th}$ step. We will also
use subscripts on non-trivial one-qubit operators to indicate which
qubit they are acting on. For example, $X$ acting on the third qubit
is written $X_3\equiv I\otimes I \otimes X \otimes I\otimes\cdots$.

We now use $H^{(1)}$ to simulate $H^{(2)}=Z_1 H^{(1)} Z_1 + H^{(1)}$.
Noting that $ZZZ=Z$, $ZXZ = -X$ and $ZYZ=-Y$, we see that the term
$H_{\alpha_Z}$ is replaced in $H^{(2)}$ by $2H_{\alpha_Z}$, but any
term in $H^{(1)}$ which has $X$ or $Y$ acting on the first qubit is
cancelled out. We then simulate $H^{(3)}=Z_2 H^{(2)} Z_2 + H^{(2)}$
and so on until we obtain $H^{(n+1)} = Z_n H^{(n)}Z_n +H^{(n)}$. In
this Hamiltonian, $H_{\alpha_Z}$ has been replaced by
$2^nH_{\alpha_Z}$ and it consists only of terms containing $Z$ or $I$.

Now we wish to remove all interactions that act on qubits outside of
$S_\alpha$. We do this by conjugating with $X$ on each of these
qubits. If $H_{\alpha_Z}$ has an $I$ acting on the $q^\text{th}$
qubit, then we simulate $H^{(n+2)}= X_{q} H^{(n+1)}X_{q} + H^{(n+1)}$.
This takes $2^nH_{\alpha_Z}$ to $2^{n+1}H_{\alpha_Z}$ and cancels any
terms that have a $Z$ acting on the $q^\text{th}$ qubit.  Let
$k\equiv|S_\alpha|$.  Since there are $n-k$ qubits upon which
$H_{\alpha_Z}$ doesn't act, if we perform this style of simulation
$n-k$ times on different qubits, then we will have removed all
interactions acting on qubits outside of $S_\alpha$. So the simulated
Hamiltonian $H^{(2n-k+1)}$ has interactions that only act on subsets
of $S_\alpha$, and $H_{\alpha_Z}$ has become $2^{2n-k}H_{\alpha_Z}$.

At this point, we wish to eliminate all remaining terms except
$H_{\alpha_Z}$.  Denote one of these undesirable terms by
$H_{\beta_Z}$.  We know that $H_{\beta_Z}$ couples a set of qubits
$S_\beta$ that is strictly contained in $S_\alpha$.  Thus, there must
be some qubit $q$ which is in $S_\alpha$ and not in $S_\beta$.  Noting
that $X_pX_q$ commutes with $Z_pZ_q$ but anticommutes with $Z_pI_q$
for $p\neq q$, we see that conjugating by $X_pX_q$ leaves
$H_{\alpha_Z}$ invariant but takes $H_{\beta_Z}$ to $-H_{\beta_Z}$ if
we choose $p$ to be in $S_\beta$.  Thus, if we simulate $H^{(2n-k+2)}=
X_{p}X_{q} H^{(2n-k+1)} X_{p} X_{q} +H^{(2n-k+1)}$ we eliminate
$H_{\beta_Z}$.  Iterating this procedure for every combination of $p$
and $q$ in $S_\alpha$, we are can eliminate every remaining undesirable
term. There are $\left(%
\begin{smallmatrix}
  k \\
  2 \\
\end{smallmatrix}%
\right)$ such possible combinations, so we finally obtain
\begin{equation} \label{eq:term_isolation}
H^{\left(2n-k+1+\left(%
\begin{smallmatrix}
  k\\
  2\\
\end{smallmatrix}%
\right)\right)} =2^{2n-k+\left(%
\begin{smallmatrix}
  k\\
  2\\
\end{smallmatrix}%
\right)}H_{\alpha_Z}.
\end{equation}

Summarizing, we can isolate any term in $H$, and thus can simulate all
elements of $C_\alpha$ for every $H_\alpha$ appearing in $H$, as well
as all linear combinations of the elements of $C_\alpha$.

We note in passing a group-theoretic
interpretation~\cite{Viola99a,Zanardi99a} of the term isolation
procedure described above.  A particular term $H_\alpha$ that we wish
to isolate forms, along with the identity, an order two group
${\mathcal G}$ which is a subgroup of the full Pauli group ${\mathcal
  P}$ on our system.  In particular $H_\alpha$ and $I$ are a
representation of this subgroup ${\mathcal G}$.  Denote the commutant
of ${\mathcal G}$ in ${\mathcal P}$ (the set of all group elements in
${\mathcal P}$ which commute with ${\mathcal G}$) as ${\mathcal
  G}^\prime$.  If we map the elements of ${\mathcal G}^\prime$ to
Pauli operators, then we have a faithful representation of ${\mathcal
  G}^\prime$. Let us denote the elements of this representation by
$D(g')$, where $g' \in {\mathcal G}$.  From Schur's lemma it then
follows that by averaging over all elements in this representation of
${\mathcal G}^\prime$ we obtain only elements in the representation of
${\mathcal G}$ given by $H_\alpha$ and $I$:
\begin{equation}
{1 \over |{\mathcal G}^\prime|} \sum_{g^\prime \in {\mathcal G}^\prime} D(g^\prime) H
D(g^{\prime})^{\dagger} = a I + b H_\alpha
\end{equation}
for some constants $a$ and $b$.
Thus we can isolate a term by performing the appropriate group average
over the commutant subgroup.  In fact, since ${\mathcal G}$ is
abelian, we can average over elements in ${\mathcal G}^\prime$ which
are not in ${\mathcal G}$.  The term isolation procedure we described
above is a concrete realization of this group average.

%
%

%
%
\subsection{Commutator restrictions}
\label{subsec:commutatorrestrictions}

We now turn to the simulation of Hamiltonians using commutators,
focusing on the possible forms of Hamiltonians simulated in this
manner.  The restrictions we obtain will be vital for the results of
the next section.

Consider two different elements of the coupling set $C_\alpha$,
$H_\alpha$ and $H_{\alpha'}$.  It is straightforward to verify from
the commutation relations for the Pauli matrices that \emph{the
commutator $[H_\alpha,H_{\alpha'}]$ is non-zero if and only if there
is an odd number of locations in $S_{\alpha}$ where the couplings
$H_\alpha$ and $H_{\alpha'}$ differ}.  For example, consider the
commutator of $H_\alpha = X\otimes X\otimes X$ and $H_{\alpha'} =
Y\otimes X \otimes Y$, both of which couple the same qubits:
\begin{equation} \begin{split}
i[H_\alpha,H_{\alpha'}]=&i(XY\otimes I\otimes XY-YX\otimes I\otimes
YX)\\ =&i(XY\otimes I\otimes XY-XY\otimes I\otimes XY)=0.
\end{split} \end{equation}
We see that in this case an even number --- two --- of the qubits are
acted on by different Paulis, and so the commutator is zero. However
if $H_{\alpha'}=Y\otimes X\otimes X$, the commutator is non-zero,
indeed it is $i[H_\alpha,H_{\alpha'}]=-2Z\otimes I\otimes I$. We see
from this example that given an initial coupling, we can generate
terms which couple a different set of qubits.

Suppose now that $C_\alpha$ and $C_\beta$ are two different coupling
sets. If they act on nonoverlapping sets of qubits, then any
commutator between an element of $C_\alpha$ and an element of
$C_\beta$ will always be zero. The commutators between two particular
Hamiltonians $H_\alpha$ and $H_\beta$ depend only on their actions on
qubits in the intersection of the sets $S_\alpha$ and $S_\beta$.
Recall that the commutator of two terms is nonzero if there is an odd
number of pairs that disagree.  Thus $H_\alpha$ and $H_\beta$ must
differ on an odd number of qubits from the intersection of $S_\alpha$
and $S_\beta$ in order to have a nonzero commutator.  To summarize, a
commutator of Hamiltonians from $C_\alpha$ and $C_\beta$ can generate
a Hamiltonian coupling a set of qubits $S_\gamma$ precisely when
$S_\gamma$ is in $S_\alpha\cup S_\beta$, $S_\gamma$ contains an odd
number of qubits from $S_\alpha\cap S_\beta$, and $S_{\gamma}$
contains all the qubits from both $S_\alpha$ and $S_\beta$ which are
not in $S_\alpha\cap S_\beta$.

This becomes clearer with an example.  Suppose $H_\alpha=X\otimes
X\otimes X\otimes X$ and $H_\beta=Z\otimes X\otimes X\otimes I$.
We find that the commutator is $i[H_\alpha,H_\beta]=-2Y\otimes
I\otimes I\otimes X$. So we see that we have simulated a two-qubit
entangling Hamiltonian, with a four- and a three-qubit coupling.

Combining all of our results about simulation so far, we see that
given $H=\sum_\alpha H_\alpha$ and arbitrary one-qubit unitaries, we
can isolate any term $H_\alpha$.  This can then be used to simulate
any Hamiltonian coupling the same set of qubits.  All of these terms
can be combined in arbitrary linear combinations.  Finally, we can use
pairs of couplings $C_\alpha$ and $C_\beta$ to simulate a coupling on
a different set of qubits $S_\gamma$ when the conditions stated above
hold.  The coupling $C_\gamma$ can be added to our repertoire of
simulatable couplings, and can be used in turn to generate new
couplings.  How many different couplings are there?  Since we assume
that there is a finite number of qubits $n$ there are no more than
$2^n$ different coupling sets, so the process of generating new
couplings and adding them to the repertoire must terminate.

Once all of the simulatable couplings have been enumerated, the
complete set of simulatable Hamiltonians consists of those whose Pauli
operator expansion contains only terms that belong to one of the
simulatable couplings.  Of course, this procedure of exhaustive
enumeration for determining which Hamiltonians are simulatable is not
especially efficient or insightful.  In the next section we provide a
surprisingly simple procedure that enables us to determine when a
Hamiltonian is universal.

\section{Structure of simulated couplings and dynamic universality}
\label{sec:universality}

%
%
In the previous section we identified various simulation methods for
Hamiltonians acting on qubits. In this section we use these methods to
classify which many-qubit Hamiltonians are universal given local
unitary operations, assuming throughout that the Hamiltonians under
consideration are entangling.  In order to find this classification we
use properties of the Pauli operator expansion of a qubit Hamiltonian
as given by equation (\ref{eq:op_exp}). In particular, we will see
that the parities of the couplings in this expansion determine whether
or not $H$ is universal.

We know from subsection \ref{subsec:termisolation} that given
$H=\sum_\alpha H_\alpha$ and local unitaries we can simulate any
particular coupling term $H_\alpha$. If we had as our base set of
operations each $H_\alpha$ in the expansion of $H$, and local
unitaries, we could simulate $H$. Thus, having $H$ and local unitaries
is equivalent to having $\{H_\alpha\}$ and local unitaries.  In this
section we focus on the simulating capacity of particular coupling
terms $H_\alpha$ as this will be sufficient for analyzing the
universality of $H$.

In section \ref{sec:background} we defined the coupling set $C_\alpha$
as being the complete set of Pauli product Hamiltonians that couple
the set of qubits $S_\alpha$. We also noted that a single element of
$C_\alpha$ and arbitrary one-qubit unitary control generates all
elements of $C_\alpha$. In this section we will often use the coupling
set $C_\alpha$ and any single element of that set interchangeably.  For
convenience, we write $LU$ to represent all local unitaries, that is,
products of one-qubit unitaries.  We also define the \emph{parity} of
a Pauli-product Hamiltonian $H_\alpha$ to be odd if it acts
non-trivially on an odd number of qubits, or, equivalently, if
$S_\alpha$ contains an odd number qubits.  Otherwise, we say that
$H_\alpha$ has even parity.  We will see that dynamic universality of
$H$ is completely determined by the parities of the terms in $H$.

For convenience, we restate a previous result that we will use
frequently.
%
%
\begin{theorem}[Bipartite Hamiltonian
theorem~\cite{Dodd02a,Wocjan02a,Bennett01a,Leung01a,Dur01a}]
\label{bipartite_ham}
Suppose $H$, acting on $n$ qubits, has only one- and two-qubit terms
in its Pauli-product expansion, and that $H$ is entangling, i.e., all
$n$ qubits are connected, possibly indirectly, by the terms in $H$.
Then $H$, together with local unitaries, is universal for quantum
computation on $n$ qubits.
\end{theorem}

We now prove a series of lemmas leading to our first new theorem.

%
%
\begin{lemma}
\label{lemma1} If Hamiltonians $H_\alpha$ and $H_\beta$ act on
sets of qubits $S_\alpha$ and $S_\beta$ such that $S_\beta \subset
S_\alpha$ and $|S_\alpha|=|S_\beta|+1$ then the set
$\{H_\alpha,H_\beta,LU\}$ is universal on $S_\alpha$.
\end{lemma}

Before giving the proof, consider the example $H_\alpha=X\otimes
X\otimes X\otimes X$ and $H_\beta=Y\otimes X\otimes X\otimes I$.
Their commutator is
\begin{equation}
i[H_\alpha,H_\beta]= -2Z\otimes I\otimes I \otimes X
\end{equation}
which means that we can simulate arbitrary couplings between qubits 1
and 4. On the other hand, the commutator of $X\otimes Y\otimes
X\otimes I\in C_\beta$ with $H_\alpha$ generates a coupling between
qubits 2 and 4.  Similarly, we can couple qubits 3 and 4.  Using
Theorem~\ref{bipartite_ham} we see that $\{H_\alpha,H_\beta,LU\}$ is
universal on the set of qubits $S_\alpha$.

\textbf{Proof:} Assume without loss of generality that the qubits are
numbered so that the first $n-1$ of them are in $S_\beta$.  Then
$X^{\otimes n}\in C_\alpha$ and $X^{\otimes n-2}\otimes Y\otimes I\in
C_\beta$, where $Y$ acts on the $(n-1)^\text{th}$ qubit. We can
simulate the commutator of these Hamiltonians
\begin{equation}
i[X^{\otimes n},X^{\otimes n-2}\otimes Y\otimes I]=-2I^{\otimes
n-2}\otimes Z\otimes X.
\end{equation}
and thus we can couple the $(n-1)^\text{th}$ and $n$th qubits. We can
perform similar simulations where $Y$ acts on each qubit in the range
1 to $n-1$.  This generates two-qubit couplings connecting all of
$S_\alpha$, and Theorem~\ref{bipartite_ham} implies that $\{H_\alpha,
H_\beta,LU\}$ is universal on $S_\alpha$. $\Box$

%
%
\begin{lemma}
\label{lemma2} If a Hamiltonian $H_\alpha$ has even parity
then $\{H_\alpha, LU\}$ is universal on $S_\alpha$.
\end{lemma}

\textbf{Proof:} This result follows almost immediately from
Lemma~\ref{lemma1}.  Let $n = |S_{\alpha}|$.  Notationally, it will be
convenient to omit qubits outside the set $S_{\alpha}$ in the
following; in all cases there is an implied identity action on the
omitted qubits.  Since $X^{\otimes n}\in C_\alpha$ and $Y^{\otimes
  n-1}\otimes X\in C_\alpha$, we can simulate the commutator
\begin{equation}
i[X^{\otimes n},Y^{\otimes n-1}\otimes X]=2i(iZ)^{\otimes n-1}\otimes
I
\end{equation}
as $n-1$ is an odd number. Since $2i(iZ)^{\otimes n-1}\otimes I$ acts
on $n-1$ qubits, Lemma 1 allows us to conclude that $\{H_\alpha,LU\}$
is universal on $S_\alpha$. $\Box$

\begin{corollary}
\label{corollary1} If $H_\alpha$ and $H_\beta$ both have even parity
and $S_\alpha\cap S_\beta \neq \O$, then $\{H_\alpha,H_\beta,LU\}$ is
universal on $S_\alpha\cup S_\beta$.
\end{corollary}

\textbf{Proof:} Lemma~\ref{lemma2} tells us that $H_\alpha$ is
universal on $S_\alpha$ and $H_\beta$ is universal on $S_\beta$.
Since they have a non-empty intersection, they can simulate a set of
two-qubit Hamiltonians connecting every qubit in $S_\alpha \cup
S_\beta$ which implies that $\{H_\alpha,H_\beta,LU\}$ is universal on
$S_\alpha \cup S_\beta$. $\Box$

%
%
\begin{lemma}
\label{lemma3} If $H_\alpha$ has odd parity, $n$, $H_\beta$ has even
parity, and $S_\beta \subset S_\alpha$, then $\{H_\alpha,H_\beta,LU\}$
is universal on $S_\alpha$.
\end{lemma}

Consider the following example. Let $H_\alpha = X^{\otimes 5}$ and
$H_\beta = I^{\otimes 3}\otimes Y\otimes X$. The commutator of these
Hamiltonians is a four-qubit coupling, and therefore, by Lemma
\ref{lemma1}, they are universal on $S_\alpha$:
\begin{equation}
i[H_\alpha,H_\beta]=-2X\otimes X\otimes X\otimes Z\otimes I
\end{equation}
Let's now generalize this example to prove Lemma~\ref{lemma3}.

\textbf{Proof:} Using $H_\alpha$ and $LU$ we can simulate the
Hamiltonian $X^{\otimes n}$, and by Lemma~\ref{lemma2} $H_\beta$ and
$LU$ can be used to simulate the Hamiltonian $I^{\otimes n-2}\otimes
Y\otimes X$ (assuming we number the qubits so that the $n^\text{th}$
and $(n-1)^\text{th}$ qubits are in $S_\alpha$).  Then we can simulate
the Hamiltonian
\begin{equation}
i[X^{\otimes n},I^{\otimes n-2}\otimes Y\otimes X] =-2X^{\otimes
n-2}\otimes Z\otimes I,
\end{equation}
which acts on $n-1$ qubits in $S_\alpha$.  Therefore,
Lemma~\ref{lemma1} implies that $\{H_\alpha,H_\beta,LU\}$ is universal
on $S_\alpha$. $\Box$

\begin{corollary}
\label{corollary2} If $H_\alpha$ has odd parity and we have a
universal set of gates $U_\beta$ acting on a set of qubits $S_\beta$
such that $S_\beta\subset S_\alpha$ and $|S_\beta| > 1$, then the set
$\{H_\alpha,U_\beta,LU\}$ is universal on $S_\alpha$.
\end{corollary}

A simple example of such a set of universal gates on three qubits is
$\{X^{\otimes 3},\text{\sc{CNOT}}\otimes I,LU\}$.

\textbf{Proof:} If we have a universal set of gates $U_\beta$ on
$S_\beta$, then it is possible to simulate a unitary operator
equivalent to a Hamiltonian evolution by $H_\gamma$ acting on an even
number of qubits $|S_\gamma|$ with $1<|S_\gamma|\leq |S_\beta|$. Then
by Lemma \ref{lemma3} the corollary is true. $\Box$

%
%

\begin{lemma}
\label{lemma4} If $H_\alpha$ has even parity, $n$, $H_\beta$ has
odd parity, $m$, and $S_\alpha\cap S_\beta \neq \O$, then
$\{H_\alpha,H_\beta,LU\}$ is universal on $S_\alpha\cup S_\beta$.
\end{lemma}

\textbf{Proof:} First, consider the case where $|S_\alpha\cap
S_\beta|=1$. Label the qubits so that $S_\alpha$ contains the first
$n$ qubits from the left and $S_\beta$ contains the first $m$ qubits
from the right. Thus $X^{\otimes n}\otimes I^{\otimes m-1}\in
C_\alpha$ and $I^{\otimes n-1}\otimes Y^{\otimes m}\in C_\beta$. We
can simulate the commutator
\begin{equation}
i[X^{\otimes n}\otimes I^{\otimes m-1},I^{\otimes n-1}\otimes
Y^{\otimes m}] =-2X^{\otimes n-1}\otimes Z \otimes Y^{\otimes m-1}.
\end{equation}
Now, this commutator acts on $|S_\alpha \cup S_\beta|=m+n-1$ qubits
which is an even number, and therefore by Lemma~\ref{lemma2} this
Hamiltonian and $LU$ are universal on these qubits.

The case $|S_\alpha \cap S_\beta| > 1$ is even simpler to prove.
Since $n$ is even, $H_\alpha$ is universal on $S_\alpha$ by Lemma
\ref{lemma2}. Now, when $|S_\alpha\cap S_\beta|>1$ we have a universal
set of gates acting on $S_\alpha\cap S_\beta$, so
Corollary~\ref{corollary2} proves that we have a universal set on
$S_\beta$, and therefore we have a universal set on $S_\alpha \cup
S_\beta$.  $\Box$

%
%
\begin{lemma}
\label{lemma-extra}
If we have a universal set of gates $U_{\alpha}$ acting on $S_\alpha$
such that $|S_\alpha| \geq 2$, and $H_\beta$ acting on $S_\beta$ such
that $S_\alpha \cap S_\beta \neq \O$, then
$\{U_{\alpha},H_{\beta},LU \}$ is universal on $S_\alpha\cup S_\beta$.
\end{lemma}

\textbf{Proof:} The proof follows simply by considering the four
possible parity combinations for $|S_{\alpha}|$ and $|S_{\beta}|$:
\begin{itemize}
\item \textbf{Case:} $|S_{\alpha}|$ even, $|S_{\beta}|$ odd.  The
  result follows from Lemma~\ref{lemma4}.
\item \textbf{Case:} $|S_{\alpha}|$ even, $|S_{\beta}|$ even.  The
  result follows from Corollary~\ref{corollary1}.
\item \textbf{Case:} $|S_{\alpha}|$ odd, $|S_{\beta}|$ even.  The
  result follows from Lemma~\ref{lemma4}, with the roles of $\alpha$
  and $\beta$ interchanged.
\item \textbf{Case:} $|S_{\alpha}|$ odd, $|S_{\beta}|$ odd.  Pick an
  even parity subset $S_{\rm ev}$ of $S_{\alpha}$ which has a
  non-trivial overlap with $S_{\beta}$.  Since we have universality on
  $S_{\alpha}$ we must also have universality on $S_{\rm ev}$.
  Lemma~\ref{lemma4} therefore implies universality on $S_{\rm ev}
  \cup S_{\beta}$.  Universality on $S_{\alpha} \cup S_{\beta}$ now
  follows from Theorem~\ref{bipartite_ham}. $\Box$
\end{itemize}

Lemma~\ref{lemma-extra} gives us a composition rule for determining
whether or not a Hamiltonian is universal on a given set of qubits. It
tells us that if we have a set qubits coupled by even-parity coupling
terms (and thus a universal set of gates on the set), then any other
qubits that they are connected to, even indirectly, will also have a
universal set defined on them. We will see from the following lemma
that this is the only way a universal set of gates can be derived.

%
%
\begin{lemma}
\label{lemma5} If $H_\alpha$ and $H_\beta$ both have odd parity, then
their commutator $H_\gamma=i[H_\alpha,H_\beta]$ is either $0$ or it
has odd parity.
\end{lemma}

\textbf{Proof:} If $S_\alpha\cap S_\beta=\O$ then
$i[H_\alpha,H_\beta]=0$, and the lemma is trivially true. In the case
where $S_\alpha\cap S_\beta\neq\O$ let $u=|S_\alpha\cap
S_\beta|$. Now, $H_\gamma \neq 0$ only when $|S_\gamma\cap
(S_\alpha\cap S_\beta)|=d$ is odd. By definition we find that
$|S_\gamma| = |S_\alpha|+|S_\beta|-2u+d$. Since $|S_\alpha|+|S_\beta|$
and $2u$ are even, but $d$ is odd, $|S_\gamma|$ must be odd for any
non-zero $H_\gamma$. $\Box$

The conclusions of these lemmas can be succinctly expressed in the
following theorem.

\begin{theorem}
\label{theorem1} An entangling Hamiltonian acting on $n$ qubits,
together with local unitary operations, is dynamically universal if
and only if the Pauli operator expansion of the Hamiltonian contains a
term with an even number of entries.
\end{theorem}

\textbf{Proof:} From Lemma~\ref{lemma5} we know that an entangling
Hamiltonian whose Pauli operator expansion contains only odd-parity
terms, together with local unitaries, can only generate other
odd-parity Hamiltonians, which shows that it is not universal.
Conversely, from Lemmas~\ref{lemma2} and~\ref{lemma5}, we see
immediately that an entangling Hamiltonian with at least one
even-parity term is universal. $\Box$

We now know that in order for a general $n$-qubit Hamiltonian to be
universal when aided by one-qubit unitaries, it must fulfill two
conditions. The first condition is that the Hamiltonian must be
entangling, in the sense explained in the introduction, i.e., all $n$
qubits must be connected, either directly or indirectly, by coupling
terms in the Hamiltonian. The second condition is that at least one of
the coupling terms must have even parity. Thus, our result reduces the
problem of determining when a Hamiltonian is universal to that of
counting the parity of terms in its Pauli operator expansion.

Let's look at a couple of examples.  It was suggested in the preprint
version of~\cite{Dodd02a} (but not the published version) that it
might be possible to construct many-body Hamiltonians which are not
universal by using results from the theory of entanglement.  For
example, the GHZ state $|GHZ\rangle = (|000\rangle+|111\rangle)/\sqrt
2$ and $W$-state $(|001\rangle+|010\rangle+|100\rangle)/\sqrt 3$ are
distinct types of entanglement which cannot be
interconverted~\cite{Dur00b}, even stochastically, by local operations
and classical communication.  This led~\cite{Dodd02a} to conjecture
that Hamiltonians such as $H_{GHZ} = |GHZ\rangle \langle GHZ|$ and
$H_{GHZ'} = |GHZ\rangle \langle 000|+ |000\rangle \langle GHZ|$ are
not universal, when assisted by local unitaries.  This conjecture
turns out to be incorrect.  Expanding in the Pauli basis and omitting
$\otimes$ for brevity, we obtain:
\begin{eqnarray}
H_{GHZ} & \propto & III+ZZI+ZIZ+IZZ- XYY \nonumber \\
 & & -YXY-YYX \\
H_{GHZ'} & \propto & III+ZII+IZI+IIZ+ZZI \nonumber \\
 & & +ZIZ+IZZ+ZZZ-XXX.
\end{eqnarray}
In both cases, we simply check that each qubit is coupled by one of
the two- or three-qubit terms, so the Hamiltonian is entangling, and
note that there are terms with even parity, so by
Theorem~\ref{theorem1} both these Hamiltonians are universal when
assisted by local unitaries.


%
%

\section{Hamiltonians with all-odd parity}
\label{sec:odd}

%
%
In the previous section we showed that the only non-universal
entangling Hamiltonians are the odd Hamiltonians, i.e., those whose
couplings all act on an odd number of qubits.  In this section we
study what dynamical operations can be achieved using such
Hamiltonians, together with local unitary operations.  We prove two
main results.

The first result is that an odd entangling Hamiltonian can be used to
simulate any other odd Hamiltonian on the system of $n$ qubits.  This
result, in combination with the results of the previous section, shows
that there are essentially only two types of entangling Hamiltonian on
$n$ qubits: Hamiltonians that are odd, and those that are not.  Within
these two classes all the Hamiltonians are essentially
inter-convertible, in the sense that any one can be used to simulate
the other.  Furthermore, the entangling Hamiltonians that are not odd
are intrinsically more powerful than the odd Hamiltonians, since they
can be used to simulate any odd Hamiltonian, but not vice versa.

The second result is to show that odd Hamiltonians can be made
universal, by using an appropriate \emph{logical basis} of qubits for
our system, similar to the ideas used in quantum error-correction.  In
particular, we show that such an interaction on $n$ qubits is
universal on a set of $n-1$ logical qubits.  In fact, we will see that
the encoding is as simple as it could be: each of the $n-1$ logical
qubits corresponds directly to one of the original $n$ qubits, while
the single qubit left over is not used.

%
%
To obtain our results we first need a simple lemma allowing us to use
an odd Hamiltonian coupling a set of qubits to generate odd
Hamiltonians acting on a subset of those qubits.

%
%
%

\begin{lemma}
\label{lemma7} If $H_\alpha$ has odd parity then we can simulate any
other odd-parity Hamiltonian $H_\beta$ provided $S_\beta\subseteq
S_\alpha$.
\end{lemma}

\textbf{Proof:} Let $|S_\alpha|\equiv2m+1$.  We prove this lemma using
induction on $m$. The lemma is trivially true for the case $m=1$ (that
is, $|S_\alpha|=3$). Now, we make the inductive assumption that the
lemma holds for the $m^\text{th}$ case and prove that it holds for the
$(m+1)^\text{th}$ case.  Given $H_{\alpha'}$ acting on a set of
$|S_{\alpha'}|=2m+3$ qubits, we need only show that we can simulate a
Hamiltonian acting on any subset of $S_{\alpha'}$ containing $2m+1$
qubits.  By assumption, we can simulate the $(2m+3)$-qubit
Hamiltonians $X^{\otimes2m+3}$ and $X^{\otimes 2}\otimes Y^{2m+1}$ and
thus their commutator
\begin{equation}
i[X^{\otimes2m+3},X^{\otimes 2}\otimes Y^{2m+1}] = -2iI\otimes I
\otimes (iZ)^{\otimes 2m+1}.
\end{equation}
Therefore we can simulate a Hamiltonian acting on the final $2m+1$
qubits of $S_\alpha$.  Similarly, we can simulate a Hamiltonian acting
on any subset of $S_\alpha$ containing $2m+1$ qubits, which proves the
inductive hypothesis and thus the lemma. $\Box$

It turns out that both the main results of this section are
corollaries of this lemma and a second result that is motivated by the
following example.  Suppose we have the ability to simulate the odd
Hamiltonian $H=ZZZII+IIZZZ$ (where we have omitted both subscripts and
tensor products).  We already know that we can therefore simulate
$ZZZII$ and $IIZZZ$ separately.  Suppose further that we wish to
simulate another odd Hamiltonian, $ZIIZZ$.  We can see that this is
possible, using our intuition about non-odd Hamiltonians, by
``isolating'' one of the qubits, say the fifth qubit, and considering
the Hamiltonians that we can simulate on the first four qubits.  So,
let's alter our example and give ourselves the ability to to perform
$ZZZI$ and $IIZZ$ on the first four qubits, and we will attempt to
simulate $ZIIZ$. Notice that now we have an odd and an even term that
connect the first four qubits, so by Theorem~\ref{theorem1}, there
must be a sequence of commutators and linear combinations that allow
us to simulate $ZIIZ$.  Here is such a sequence, where in each step we
generate a new coupling to add to our set of allowed couplings:
\begin{equation} \begin{split}
i[ZZYI,IIXZ]&=2ZZZZ,\\
i[XZXX,YZYY]&=2ZIZZ,\\
i[ZIZY,IIZX]&=2ZIIZ.
\end{split} \end{equation}
Now, if we consider the original problem on five qubits, we see
that the odd parity restriction shows that this procedure
generates the desired coupling $ZIIZZ$:
\begin{equation} \begin{split}
i[ZZYII,IIXZZ]&=2ZZZZZ,\\
i[XZXXZ,YZYYZ]&=2ZIZZI,\\
i[ZIZYI,IIZXZ]&=2ZIIZZ.
\end{split} \end{equation}

In a similar vein, suppose that we wish to simulate an even-parity
coupling $ZIIZI$.  We know that this is not possible directly
(otherwise $H$ would be dynamically universal), but it is possible
using a very simple encoding.  If we place the fifth qubit in the
$Z$-eigenstate $|0\rangle$, then our procedure above for simulating
$ZIIZZ$ allows us to simulate $ZIIZ$ on the first four qubits.  Thus,
$H$ is dynamically universal on the first four qubits.

Our main results of this section generalize these two examples.  Both
examples rely crucially on the fact that there was a qubit that could
be \emph{isolated}, that is, acted on by only a single member of the
set of couplings that we used to do our simulation.  For example,
suppose we tried to use the same approach to show that $H$ can
simulate $ZIIZ$ on four of the five qubits by placing the third qubit
in the state $|0\rangle$.  Then the two couplings that we have at our
disposal on the remaining four qubits are $ZZII$ and $IIZZ$, which do
not connect the four qubits, and thus cannot be universal on them.  We
formalize this intuition in the following lemma.

\begin{lemma}
Let $H=\sum H_\alpha$ be an odd entangling $n$-qubit Hamiltonian, and
$\mathcal{H}=\{H_\alpha\}$ be the set of all terms in $H$.  Then there
exists a set $\mathcal{M}\subseteq\mathcal{H}$ that (a) connects all
$n$ qubits in such a way that (b) at least one of the qubits is only
acted on by a single element of $\mathcal{M}$.  We call such a qubit
an \emph{isolated qubit}, and $\mathcal{M}$ an \emph{isolating set}
for that qubit.
\end{lemma}

\textbf{Proof:} We prove this lemma by giving a constructive procedure
to generate an appropriate set $\mathcal M$.  For convenience, define
$n_\alpha\equiv|S_\alpha|$ for all $\alpha$.
\begin{enumerate}
\item Choose a term $H_1$ from the set $\mathcal H$.  Without loss of
generality, we may number the qubits so that it acts on the first
$n_1$ qubits.  Add $H_1$ to the set $\mathcal M$.
\item Search for a second term in the set $\mathcal H$ that overlaps
with $H_1$ and also acts on at least one qubit outside of $S_1$.  If
there is no such term, then $H_1$ must couple all of the qubits, in
which case $\mathcal{M}=\{H_1\}$ satisfies the conditions above and we
are done.
\item Otherwise, choose such a term and call it $H_{2'}$.  Define
$n_{S_1\cap S_{2'}}$ to be the number of qubits in $S_1\cap S_{2'}$.
Without loss of generality, we may assume that these qubits are strung
out in a line, with the $n_1$ left-most qubits in $S_1$ and the
$n_{2'}$ right-most qubits in $S_{2'}$, and the $n_{S_1\cap S_{2'}}$
overlapping qubits in the middle.
\item If $n_{S_1\cap S_{2'}}$ is odd, then use $H_{2'}$ to simulate a
Hamiltonian $H_2$ that acts on qubits $n_1,...,n_1+n_{2'}-n_{S_1\cap
S_{2'}}$ (where we number from the left, starting at 1).  $H_2$ acts
on the right-most $n_2\equiv n_{2'}-n_{S_1\cap S_{2'}}+1$ qubits,
overlapping with $H_1$ on just a single qubit (the $n_1$th qubit).
This is possible since $n_2$ is odd and $S_2\subseteq S_{2'}$,
satisfying the conditions of Lemma~\ref{lemma7}.  Add $H_2$ to
$\mathcal M$.
\item On the other hand, if $n_{S_1\cap S_{2'}}$ is even, then we use
$H_{2'}$ to simulate a Hamiltonian $H_2$ that acts on qubits
$n_1-1,...,n_1+n_{2'}-n_{S_1\cap S_{2'}}$.  This $H_2$ acts on the
right-most $n_2\equiv n_{2'}-n_{S_1\cap S_{2'}}+2$ qubits, overlapping
with $H_1$ on just two qubits, in positions $n_1-1$ and $n_1$.  Again,
$n_2$ is odd and $S_2\subseteq S_{2'}$, satisfying the conditions of
Lemma~\ref{lemma7}.  Add $H_2$ to $\mathcal M$.
\item Now, if there are no other terms that overlap $H_2$, then the
right-most qubit in $S_2$ must be isolated since $H_2$ acts on at
least three qubits, and overlaps with $H_1$ on at most two.  If we
then add the remaining Hamiltonians from $\mathcal H$ (i.e., all
except for $H_1$ and $H_{2'}$) to $\mathcal M$, then $\mathcal M$ must
couple all $n$ qubits and contain an isolated qubit, satisfying the
conditions above, and so we are done.
\item Otherwise, repeat steps 2 to 5 to generate a Hamiltonian $H_3$
that overlaps only with $H_2$ on one or two qubits, and add it to
$\mathcal M$.  Repeat this process, adding a Hamiltonian to $\mathcal
M$ each time, until it becomes impossible to find a term that both
overlaps with the previous term and acts on at least one more qubit
than it.  When the process terminates (which must happen eventually
since there is only a finite number of qubits), the last term that was
added must contain an isolated qubit.  If $\mathcal M$ connects all
$n$ qubits then we are done, otherwise add the remaining Hamiltonians
from $\mathcal H$ to $\mathcal M$ to complete the construction.
\end{enumerate}  $\Box$

Using this lemma, we can prove our two main results.

\begin{theorem}
\label{theorem2}
Let $H$ be an odd $n$-qubit entangling Hamiltonian.  Then $H$ and $LU$
can simulate any odd Hamiltonian on the $n$ qubits.
\end{theorem}

\textbf{Proof:} Let $\mathcal{M}$ be an isolating set for a qubit,
which, without loss of generality, we may choose to be the
$n^\text{th}$ qubit.  Suppose we consider the set of couplings
$\mathcal M'$ on the first $n-1$ qubits that arises by simply taking
the couplings in $\mathcal M$, and omitting the Pauli acting on the
final qubit.  Then by construction of $\mathcal M$ we see that this
set (a) connects the first $n-1$ qubits, and (b) contains an element
that acts on an even number of these qubits, corresponding to the
element of $\mathcal M$ that couples to the isolated qubit.  By
Theorem~\ref{theorem1} it follows that $\mathcal M'$, together with
local unitaries, is universal on the first $n-1$ qubits.

Lifting back up to the full set of $n$ qubits, we see that $\mathcal
M$ must generate the set of all odd couplings on the $n$ qubits.  To
see this a little more explicitly, suppose that we wish to generate an
odd coupling $\sigma$.  Let $\sigma'$ be the corresponding coupling on
the first $n-1$ qubits.  By an appropriate sequence of commutators of
elements of $\mathcal M'$ we can generate $\sigma'$.  The
corresponding sequence of commutators in $\mathcal M$ must generate
$\sigma$, up to possible relabeling on the final qubit, which can be
accomplished via appropriate local unitaries.  $\Box$

\begin{theorem}
\label{theorem3}
Let $H$ be an odd $n$-qubit entangling Hamiltonian.  Then $H$ and $LU$
are universal on a set of $n-1$ logical qubits.
\end{theorem}

Before proving this theorem, let's consider another example.  Consider
the Hamiltonian $H_\alpha = Z\otimes Z\otimes Z$. We know from the
previous section that this Hamiltonian and local unitaries do not form
a universal set of operations on three qubits. However, if we prepare
the third qubit in the $|0\rangle$ eigenstate of $Z$ at the beginning
of the procedure, then any succession of evolutions of $H_\alpha$ and
local unitaries acting only on the first two qubits will leave the
third qubit invariant throughout the evolution. So if, for instance,
we trace over the third qubit we find that the reduced system evolves
according to $Z\otimes Z$. We know that this Hamiltonian is universal
on the reduced system, so in effect we have a universal set of gates
on the first two qubits. Essential to this example is our ability to
identify a qubit that can be prepared in a local eigenstate of the
Hamiltonian that entangles the other qubits.

\textbf{Proof:} The proof is simply to number the qubits $1,\ldots,n$,
and to prepare the $n^\text{th}$ qubit in a fixed state $|0\rangle$.
Suppose now that we wish to simulate an arbitrary Pauli $\sigma$
acting on the first $n-1$ qubits.  If $\sigma$ has odd parity, then we
can use the results of Theorem~\ref{theorem2} to simulate $\sigma$
directly.  If $\sigma$ has even parity then we use the results of
Theorem~\ref{theorem2} to simulate $\sigma \otimes Z$, which leaves
the state of the final qubit unchanged, and evolves the first $n-1$
qubits according to the Hamiltonian $\sigma$.  Thus, we can use this
construction to simulate any interaction on the first $n-1$ qubits.
$\Box$

We now understand that odd $n$-qubit Hamiltonians, in contrast to the
non-odd Hamiltonians, do not generate the algebra
$\mathfrak{su}(2^n)$.  What algebra do they generate?\footnote{This
  question was originally posed to us by Greg Kuperberg.}  It turns
out that the answer depends on whether $n$ is odd or even.

%
%
To state and prove our results, it is helpful to be a little more
precise about the various Lie algebras we are considering.  We define
$\mathfrak g$ to be the real Lie algebra generated by odd parity
Paulis acting on $n$ qubits.  More precisely, $\mathfrak{g}$ is a
real vector space whose basis elements are of the form $i \sigma$,
where $\sigma$ is an odd parity Pauli.  We have shown that this is the
relevant Lie algebra associated with an entangling Hamiltonian on $n$
qubits, plus one-qubit unitaries.  The following theorem relates
$\mathfrak g$ to the standard classification of Lie algebras:

\begin{theorem} \label{th:lie}
  If $n$ is even, then $\mathfrak g$ is isomorphic to
  $\mathfrak{so}(2^{n})$.  If $n$ is odd, then $\mathfrak g$ is
  isomorphic to $\mathfrak{sp}(2^n)$.  Furthermore the representation
  of $\mathfrak g$ provided by tensor products of odd parity Paulis on
  $n$ qubits is the fundamental (defining) representation of these Lie
  algebras.
\end{theorem}

Amusingly, the $n = 2$ case of this theorem is a well-known result from
Lie theory, the isomorphism between $\mathfrak{so}(4)$ and
$\mathfrak{su}(2) \otimes \mathfrak{su}(2)$.  This result has received
wide use in quantum information theory in a different guise --- it is
just the fact that local (special) unitary operations on two qubits
correspond to real orthogonal transformations in the so-called ``magic
basis''.

\textbf{Proof:} We consider the $n$ even case first.  Let us define an
operation $f(\sigma) \equiv (-1)^{{\rm wt}(\sigma)} \sigma$ on Pauli
matrices, where ${\rm wt}(\sigma)$ is the weight of $\sigma$.  This
operation can be extended by linearity to all matrices.  Observe that:
\begin{eqnarray}
  f(A) = Y^{\otimes n} A^T Y^{\otimes n}.
\end{eqnarray}
Then the Lie algebra $\mathfrak g$ consists of all matrices $A$ such
that:
\begin{eqnarray} \label{eq:g-defn}
  f(A) = -A, \quad {\rm and} \quad A^\dagger = -A.
\end{eqnarray}

%
%
The Lie algebra $\mathfrak{so}(2^n)$ can be defined similarly.  Recall
that the defining representation of the Lie algebra
$\mathfrak{so}(2^n)$ consists of $2^n \times 2^n$ matrices ${B}$ which
satisfy~\cite{Georgi99a,Cornwell97a}
\begin{equation} \label{eq:so-defn}
{B}^T=-{B}, \quad {\rm and} \quad {B}^\dagger=-{B},
\end{equation}
where $T$ denotes the transpose operation.

%
%
We aim to find a unitary $U$ such that $B$ satisfies
Eq.~(\ref{eq:so-defn}) if and only if $A = U B U^\dagger$ satisfies
Eq.~(\ref{eq:g-defn}).  It is easy to see that for any $U$, $A^\dagger
= -A$ if and only if $B^\dagger = -B$, so we need only find a $U$ such
that $f(A) = -A$ if and only if $B^T = -B$.  Straightforward algebraic
manipulation shows that $U = (I-iY^{\otimes n})/ \sqrt{2}$ satisfies
this requirement.

%
%
The $n$ odd case is very similar.  The defining representation of the
Lie algebra $\mathfrak{sp}\left(2^n\right)$ consists of $2^n \times
2^n$ matrices which satisfy \cite{Georgi99a,Cornwell97a}
\begin{equation} \label{eq:sp}
{J}^\dagger {B}^T {J} = -  { B}, \quad {\rm and} \quad { B}^\dagger = - { B},
\end{equation}
where $J = Y_1 \otimes I$ is the Pauli $Y$ acting on the first qubit
alone. Setting $U = I \otimes (I-iY^{\otimes n-1})/\sqrt 2$, we see
that Eq.~(\ref{eq:sp}) is equivalent to Eq.~(\ref{eq:g-defn}) if we
set $A = UB U^\dagger$.
$\Box$

%
%
An interesting consequence of Theorem~\ref{th:lie} occurs for odd
parity Paulis acting on an odd number of qubits, $n$.  Suppose we are
given some known pure state, $|\psi\rangle$.  We can then ask the
question of whether we can transform this state into any other state,
$|\phi\rangle$.  Clearly if we have control over $\mathfrak {su}(2^n)$
we can perform this task.  A theorem from the study of the
controllability of quantum systems~\cite{Schirmer02a,Schirmer02b}
shows that this task can be performed for all states $|\psi\rangle$
and $|\phi\rangle$ if and only if one has control over the Lie algebra
$\mathfrak{su}(2^n)$ or the Lie algebra $\mathfrak{sp}(2^n)$.  Thus,
while we do not have full unitary control when we have odd parity
Paulis acting on an odd number of qubits (except when $n = 1$), we can
transform any state into any other state using these operations.

\section{Efficiency}
\label{sec:efficiency}

We have examined the problem of Hamiltonian simulation across sets of
qubits that are coupled by a fixed natural Hamiltonian.  However, our
results appear to be limited by the fact that our simulation
techniques are manifestly exponential in the total number of qubits
$n$.  The problem is that our procedure for isolating a single term of
an arbitrary Hamiltonian requires on the order of $2^n$ local
unitaries to be interspersed in the evolution at different times (see
equation~(\ref{eq:term_isolation})).  This is in sharp contrast to the
case of Hamiltonians whose Pauli expansions contain only two-qudit
couplings, and simulation techniques are polynomial in the number of
qudits.  This defect could be fixed if we were to find efficient
techniques for term isolation, for all other steps in our procedure
were efficient, at least in principle.

It is not surprising that, in this very general situation, we have not
obtained efficient simulation techniques.  Suppose, for example, that
we have a family of Hamiltonians that has the property that the member
of the family that acts on $n$ qubits has a decomposition containing a
tensor product of $X$s acting on every subset of those $n$ qubits:
\begin{equation}\begin{split}
H_{(2)}&=X_1 X_2\\
H_{(3)}&=X_1 X_2+X_1 X_3+X_2 X_3+X_1 X_2 X_3\\
H_{(4)}&=X_1 X_2+\cdots+X_1 X_2 X_3+\cdots+X_1 X_2 X_3 X_4\\
&\vdots
\end{split}\end{equation}
where the subscripts in brackets indicate the number if qubits acted
on by each Hamiltonian.  These Hamiltonians are not at all natural
since every possible coupling of qubits is represented, regardless of
how ``far away'' the qubits are from one another.  With such a large
number of couplings ($2^n-n-1$ for the $n$-qubit member of the
family), it is not surprising that we have not found a method to
efficiently simulate a set of gates that is universal for quantum
computation, since it is difficult to ``turn off'' enough unwanted
interactions, while retaining computational universality, even though
these Hamiltonians are dynamically universal.  As mentioned in the
introduction, we conjecture that a generic entangling Hamiltonians
will not be universal for quantum computation.

On the other hand, it would be surprising if the following family of
Hamiltonians could not be used simulate a universal set of gates:
\begin{equation}\begin{split}
H_{(2)}&=X_1 X_2\\
H_{(3)}&=X_1 X_2+X_2 X_3+X_1 X_2 X_3\\
H_{(4)}&=X_1 X_2+X_2 X_3+X_3 X_4+X_1 X_2 X_3+X_2 X_3 X_4\\
H_{(5)}&=X_1 X_2+\cdots+X_4 X_5+X_1 X_2 X_3+\cdots+X_3 X_4 X_5\\
&\vdots
\end{split}\end{equation}
In these Hamiltonians, each qubit is directly coupled to at most four
other qubits by terms acting on at most three qubits.  For example,
qubit 3 only ever couples directly to qubits 1, 2, 4, and 5.
Furthermore, the number of coupling terms is linear in the number of
qubits --- the Hamiltonian on $n$ qubits has only $2n-3$ terms in its
decomposition.  This number is sufficiently small, by contrast with
the general case, that we might hope that it is possible to turn off
most (but not all) of these couplings in an efficient fashion. We call
a Hamiltonian (or, more precisely, a family of Hamiltonians)
\emph{$k$-local} if the maximum number of qubits coupled by any term
in its decomposition is $k$ and if the absolute values of the non-zero
coefficients multiplying each coupling in $H$ (the $h_{j_1,...,j_n}$
in the expansion of H in equation (\ref{eq:op_exp})) are bounded below
by a constant. This family of Hamiltonians is thus 3-local.

Motivated by these examples, we now describe a randomized algorithm
which shows that a single term in a $k$-local Hamiltonian can be
isolated with high probability in a number of steps that is
exponential in $k$ but polynomial in $n$.  Thus, given a family of
$k$-local Hamiltonians, for some fixed $k$, we have a procedure to
efficiently simulate a set of gates that is universal for quantum
computation.

More precisely, suppose we wish to isolate a single term $H_\alpha$ in
the expansion of $H$ using local unitaries.  We give a procedure to
use a randomly chosen set of local unitaries $\{U_j\}$ to isolate
$H_\alpha$ with failure probability bounded above by $N/2^m$, where
$N$ is an upper bound on the number of terms in $H$ and $m$ is the
number of local unitaries in the set $\{U_j\}$.

If $H$ is $k$-local, then $N$ is polynomial in $n$ --- a simple bound
on $N$ is $n^k$ --- and hence the probability of failure is polynomial
in $n$ and decreases exponentially in the number of local unitaries.
More precisely, to bound the probability of failure to be less than
$\epsilon$, it turns out to be sufficient to choose $m\geq
\log(N/\epsilon)$.  The number of timesteps required increases by a
factor of two for each extra unitary, so the number of timesteps is
bounded above by $2^m=N/\epsilon$.  Provided $N$ is a polynomial
function of $n$, as is the case if $H$ is $k$-local, the number of
timesteps is also polynomial in $n$.

\textbf{Algorithm:} To explain the algorithm, we begin by explaining
how to eliminate a single unwanted term from $H$, without worrying
about keeping our desired term.  (We will see later that a simple
modification of this procedure eliminates the unwanted term and keeps
the desired term.)  Suppose, without loss of generality, that the
unwanted term has the form $H_\beta=Z_1 Z_2\cdots Z_l I_{l+1}\cdots
I_n$.  In order to eliminate this term, we choose a set of local
unitaries $\{U_j\}$, each of which is a tensor product of $n$
unitaries each chosen independently with equal probability from the
set $\{I,X,Y,Z\}$. We then do the following conjugations:
\begin{equation} \begin{split}
H^{(1)}&=U_1 H U_1 + H,\\
H^{(2)}&=U_2 H^{(1)} U_2 + H^{(1)},\\
&\vdots\\
H^{(m)}&=U_j H^{(m-1)} U_j + H^{(m-1)}
\end{split} \end{equation}
The probability that $H^{(m)}$ still contains the unwanted term
$H_\beta$ is the probability that each $U_j$ commutes with $H_\beta$.
The probability that a particular $U_j$ commutes with $H_\beta$ is
simply the probability that the total number of $X$ and $Y$ terms in
$U_j$ that act on the first $l$ qubits is even.  So, if $l$ is 1, then
the probability that $U_j$ commutes with $H_\beta=Z_1$ is just the
probability that $U_j$ has a $Z$ or an $I$ on the first qubit, which
is $1/2$.  Similarly, if $H_\beta$ acts nontrivially on the first 2
qubits, then the probability that $U_j$ commutes with it is the
probability that its first two terms are $I_1I_2$, $Z_1Z_2$, $X_1X_2$,
$X_1Y_2$, $X_1Z_2$, $Y_1X_2$, $Y_1Y_2$, or $Y_1Z_2$, which is again
$1/2$.  It is not hard to see that this pattern holds for any choice
of $H_\beta$ --- the probability that it commutes with a randomly
chosen $U_j$ is always $1/2$.  Thus, the probability that all $m$ of
the $U_j$ commute with $H_\beta$ is simply $1/2^m$.

Now impose the constraint that we wish to keep a particular term
$H_\alpha$ while eliminating $H_\beta$ in this way. Instead of
choosing $U_j$ completely randomly, we instead generate a random
product of Paulis, and then check to see if it commutes with our
desired term, $H_\alpha$.  If it does, which happens with probability
$\frac{1}{2}$, then we add it to the set $\{U_j\}$, otherwise we
repeat the process.  A simple case analysis now shows that the
probability of a given $U_j$ in this set commuting with $H_\beta$ is
$\frac 12$.  Thus, the probability that all $m$ of the $U_j$ chosen in
this fashion commute with $H_\beta$ is again $1/2^m$.

In general there will be many terms in the expansion of $H$ that we
wish to eliminate while isolating $H_\alpha$.  The probability that
$H^{(m)}$ contains a term other than $H_\alpha$ (i.e., our procedure
has failed) is certainly no greater than the sum of the probabilities
that the procedure failed to eliminate each term $H_\beta$, that is
$N/2^m$ where $N$ is the number of terms in $H$ that must be
eliminated.  $\Box$

\section{Summary and future directions}
\label{sec:discussion}

We have examined the problem of simulating Hamiltonians using a fixed
multi-qubit Hamiltonian and local unitary operations. We have provided
a classification scheme for the simulations that are possible given
these resources. In particular, we have demonstrated that there are
only two physically distinct classes of entangling Hamiltonians up to
local unitary operations. One class, the class of odd entangling
Hamiltonians, when assisted by local unitary operations, can simulate
all odd Hamiltonians but nothing else. The other class, the class of
non-odd entangling Hamiltonians, can simulate all Hamiltonians and are
thus dynamically universal. We have also demonstrated that all odd
entangling Hamiltonians can be made universal through the use of a
simple encoding scheme.  Furthermore, the Lie algebras generated by
the odd entangling Hamiltonians together with local unitary operations
were shown to be isomorphic to either $\mathfrak{so}(2^n)$ or
$\mathfrak{sp}(2^n)$, depending on whether $n$ is even, or odd,
respectively.

In this paper we have made a distinction between sets of resources
that are universal for quantum computation and those that are
dynamically universal. This distinction has been necessary because we
can not find an efficient means to simulate the Hamiltonians used in
quantum computation with an arbitrary fixed entangling Hamiltonian and
local unitaries.  We have demonstrated that when restricted to
$k$-local Hamiltonians with lower-bounded coefficients dynamic
universality is equivalent to universality for quantum computation.
The resolution of when this equivalence holds in general would be an
interesting contribution to the study of quantum dynamics.

\acknowledgments

We thank Andrew Childs, Andrew Doherty, Charles Hill, Andrew Hines,
Greg Kuperberg and John Preskill for valuable discussions. MJB, JLD,
and MAN thank the Institute for Quantum Information for their
hospitality.  This work was supported in part by the National Science
Foundation under grant EIA--0086038.

\end{document}